\documentclass[12pt]{article}
\usepackage[cp1251]{inputenc}
\usepackage[english]{babel}
\usepackage{amsfonts}

\newtheorem{lemma}{Lemma}
\newtheorem{theorem}{Theorem}
\begin{document}


\title{On a Commutative Ring of Two Variable Differential Operators with Matrix Coefficients
\author{A.E. Mironov}}
\date{}

\maketitle
\section{Introduction}

In this work, we construct commutative rings of two variable
matrix differen\-tial operators that are isomorphic to a ring of
meromorphic functions on a rational manifold obtained from the
  ${\mathbb C}P^1\times {\mathbb C}P^1$ by identification of two
lines  with the pole on a certain rational curve.

The commutation condition for differential operators is equivalent
to a system of non-linear equations in the operators'
coefficients.  For selected operator coefficients, the commutation
equations reduce to known soliton equations such as the
Korteweg-de Vries equation, the Kadomtsev-Petviashvili equation,
the $\sin$-Gordon equation and others.

The problem of classifying commuting ordinary differential
operators was solved in  \cite{K1}. If two differential operators
$$
 L_1=\partial_x^n+u_{n-1}\partial_x^{n-1}+\dots+u_0(x),\
 L_2=\partial_x^m+v_{m-1}\partial_x^{m-1}+\dots+v_0(x)
$$
commute, then by the Burchnall-Chaundy lemma \cite{BC} there
exists a non-zero polynomial  $Q(\lambda,\mu)$ of two  commuting
variables  $\lambda$ and  $\mu$ such that
$$
 Q(L_1,L_2)=0.
$$
The smooth compactification of the curve given in ${\mathbb C}^2$
 by the equation
$$
 Q(\lambda,\mu)=0
$$
is called {\it spectral curve}.
If  $\psi$ --- is a common eigen-function, and $\lambda$ and
$\mu$ --- the corresponding eigen-values$$
 L_1\psi=\lambda\psi,\
 L_2\psi=\mu\psi,
$$
then point $P=(\lambda,\mu)$ lies on the spectral curve . In this
way, the spectral curve parametrizes the commom eigen-functions of
the operators $L_1$ and $L_2$. The function $\psi(x,P)$ (the
Baker-Akhiezer function)  has a unique essential singularity on
the spectral curve, and outside of this point it is meromorphic.
The Baker-Akhiezer function one can find uniquely from its
spectral data, i.e. from the set of poles, the essential singular
point and certain relations of the residues in the poles. In this
way, the commutative rings of differential operators correspond to
the sets of spectral data. There is no such classification for
operators depending on more than one variable. In the
multidimensional  case we have the Burchnall-Chaundy lemma
analogue proven by Krichever \cite{K2}. If the operators
$L_1,\dots,L_{n+1}$ in $n$ variables whose lead symbols have
constant coefficients, do commutate pairwise, then there exists a
polinomial $Q(\lambda_1,\dots,\lambda_{n+1})$ in the commutative
variables $\lambda_1,\dots,\lambda_{n+1}$, such that
$$
 Q(L_1,\dots,L_{n+1})=0.
$$
As in the one-dimensional case, the manifold given in ${\mathbb
C}^{n+1}$ by equation $Q(\lambda_1,\dots,\lambda_{n+1})=0$
parametrizes the common eigen-functions of operators
$$
 L_1\psi=\lambda_1\psi,\dots, L_{n+1}\psi=\lambda_{n+1}\psi.
$$

The question of how to correctly compactify this spectral manifold
and to correctly assign the spectral data for unique recovery of
the Baker-Akhiezer function, and, consequently, the commutative
ring of differential operators, remains entirely open.

 In \cite{P}, \cite{ZhO}, the authors find a formal generalization of
the Krichever construction in the two-dimensional case.

There are several approaches to the construction of
multidimensional commutative operators (\cite{Pr} offers insight
into some of them). Sometimes, for operators of a certain kind the
commutation equation can be integrated (see e.g. \cite{BSh}). In
\cite{CV}, \cite{CFV} (see also the references in these papers )
the authors propose a method to recover the Baker-Akhiezer
function by the affine part of certain rational manifolds. The
enumerated examples are very interesting since the commutative
rings contain Schr\"odinger operators.

Nakayashiki \cite{N1} (see also \cite{N2}) found spectral data
corresponding to commuting differential operators in $g$ variables
with matrix coefficients (the matrix size being $g!\times g!$) ,
with principle polarized Abelian manifolds with a non-singular
theta-divisor serving as spectral manifolds. For $g=2$ these
operators have been studied in \cite{M1}, \cite{M2}. Note the main
distinction between the one-dimensional case from the
multidimentional one: Even if we could construct operators $L_1$
and $L_2$, with a sufficiently large family of common
eigen-functions, e.g. a family of parametrized points of a certain
algebraic manifold, then, in contrast to the one-dimensional case
this does not imply that operators $L_1$ and $L_2$ commute,
because the kernel of the commutator $[L_1,L_2]$  is in the
general case infinite dimensional. This difficulty is overcome in
\cite{N1} as follws. Nakayashiki constructs a module $M$ (the
Baker-Akhiezer module) above a ring of differential operators
${\cal D}_g={\cal O}[\partial_{x_1},\dots,\partial_{x_g}],$ where
${\cal O}$
 is a ring of analytic functions in variables $x_1,\dots,x_g$ in the neighborhood of
$0\in {\mathbb C}^g$, which consists of functions that depend on
$x=(x_1,\dots,x_g)$ and $P\in X$, and which possesses the
following two remarkable characteristics:

1. $M$ is a free module above ${\cal D}_g$ of rank $g!$

2. For any meromorphic function $\lambda$ on $X$ with pole on a
theta-divisor and for any function $\varphi\in M$, function
$\lambda\varphi$ also lies in $M$.

From this construction follows that there exists an imbedding of
the ring $A_{\theta}$ of meromorphic functions on $X$ with pole on
the theta-divisor into a ring of differential operators on $g$
variables with matrix coefficients. Take base
$\psi_1(x,P),\dots,\psi_{g!}(x,P)$ in the ${\cal D}_g$-module $M$.
Denote the vector function $\psi(x,P)$
$$
 (\psi_1(x,P),\dots,\psi_{g!}(x,P))^{\top}
$$
by $\psi (x,P)$. Then for any meromorphic function $\lambda (P)\in
A_{\theta}$ there exists a uniqe differential operator
$D(\lambda)$ such that

$$
 D(\lambda)\psi(x,P)=\lambda(P)\psi(x,P).
$$
Since $\lambda$ and $\mu$ do not depend on $x$, equality
$$
 D(\lambda)D(\mu)\psi=
 D(\mu)D(\lambda)\psi=D(\mu\lambda)\psi=\mu\lambda\psi,
$$
holds for any two functions $\lambda, \mu\in A_{\theta}$.

$$
 D(\lambda)D(\mu)\psi=
 D(\mu)D(\lambda)\psi=D(\mu\lambda)\psi=\mu\lambda\psi,
$$
consequently, by virtue of the differential operators' uniqueness
we have
$$
 D(\lambda)D(\mu)=
 D(\mu)D(\lambda)=D(\mu\lambda),
$$
i.e., operators $D(\lambda)$ and $D(\mu)$ commute.

Rothstein \cite{R} (see also \cite{R1}) shows that a ring of
meromorphic functions on a Fano surface with a certain fixed pole
can be imbedded into a ring of matrix differential operators.

In this work, we construct an analogue of Nakayashiki's
construction for the following spectral manifold.

Let $\Gamma$ denote a manifold derived from the ${\mathbb
C}P^1\times {\mathbb C}P^1$ identification of two straight lines
$$
 p_1\times {\mathbb C}P^1 \sim {\mathbb C}P^1\times p_2.
$$
Namely, upon assigning coordinates $(z_1:w_1,z_2:w_2)$ on
${\mathbb C}P^1\times {\mathbb C}P^1$, we identify the following
points:
$$
 (a_1:b_1,t_1:t_2)\sim (t_1:t_2,a_2:b_2),
$$
with $(a_i:b_i)$ being coordinates of $p_i$, $p_1\ne p_2$,
$(t_1:t_2)\in{\mathbb C}P^1$. Let $f(P)$ denote the following form
on ${\mathbb C}P^1\times {\mathbb C}P^1$
$$
 f(P)=\alpha z_1z_2+\beta z_1w_2+\gamma z_2w_1+\delta
 w_1w_2,\ P=(z_1:w_1,z_2:w_2)\in\Gamma,
$$
$\alpha,\beta,\gamma,\delta\in{\mathbb C},$ and suppose that
identity
$$
 f(a_1:b_1,t_1:t_2)-A f(t_1:t_2,a_2:b_2)=0,\
 A\in {\mathbb C}^*.\eqno{(1)}
$$
holds for all $(t_1:t_2)$

Equality (1) restricts the choice of constants
$a_i,b_i,\alpha,\beta,\gamma,\delta$ and means that form $f(P)$ is
a section of a line bundle on $\Gamma$.

Let $A_f$ denote a ring of meromorphic functions on $\Gamma$ with
a pole on a curve given by equation  $f(P)=0$. The main result of
this work is
\begin{theorem}
There exists an embedding
$$
 D:A_f\rightarrow Mat(2,{\cal D})
$$
of a ring of meromorphic functions $A_f$ on $\Gamma$ into the ring
$2\times 2$-matrix differential operators in the variables $x$ and
$y$.
\end{theorem}

In section 2, we introduce a Baker-Akhiezer module corresponding
to the manifold $\Gamma$. In section 3, we present theorem 2 and
show that this module is free of rank 2. Theorem 1 follows
directly from theorem 2. In section 3 we also give explicit
examples of commuting operators. It is remarkable that by
rationality of $\Gamma$, the operator coefficients are elementary
functions.

\noindent {\bf Acknowledgement}

\noindent The author would like to thank
 ESF Research Network MISGAM for partial support of the ISLAND-3 conference,
 where the main results of this work were
 obtained, and also appreciate to Oleg Chalyh and Alexander
 Veselov for useful discussions of this work.
The author is thankful to Atsushi Nakayashiki for the invitations
in Kyushu University, useful discussions of our results,
promotional interest, and important remark proposed to this work
(see paragraph 2.1). The author is also grateful to Grant-in-Aid
for Scientific Research (B) 17340048 for financial support of the
visits in Kyushu University.

\section{The Baker-Akhiezer module }
In this section we point to some obvious and necessary conditions
that should be fulfilled by the spectral data of commutative rings
of multidimensional differential operators. Also, we construct a
Baker--Akhiezer module on mani\-fold $\Gamma$.
\subsection{General construction }

As noted before, in many significant examples of commutative rings
of differen\-tial operators the common eigen-functions are
parametrized by points of a spectral manifold $X$, the ring of
operators itself being isomorphic to a ring $A_Y$ of meromorphic
functions on $X$ with pole on a hypersurface $Y\subset X$. Let the
operators have matrix size $k\times k$ coefficients, with
$\psi(x,P)=(\psi_1(x,P),\dots,\psi_k(x,P))^{\top}$ being a common
vector-eigen-function (for sake of simplicity we consider the case
of {\it rank 1} operators, i.e. where to every point of the
spectral manifold there corresponds a unique common
eigen-function) and let $x=(x_1,\dots,x_n)$, $P\in X$ be a
spectral parameter. For all examples, the unique differential
operator $D(\lambda)$ can be recovered by means of the meromorphic
function $\lambda\in A_Y$ such that $$
 D(\lambda)\psi=\lambda\psi.\eqno{(2)}
$$
Let's consider module $M$ over ${\cal D}_n$, generated by
functions $\psi_1,\dots,\psi_k$
$$
 M=\{d_1\psi_1+\dots+d_k\psi_k,\ d_i\in {\cal D}_n\}.
$$
As follows from the uniqueness of operator $D(\lambda)$, which
satisfies equality (2), the ${\cal D}_n$-module $M$ is free.
Actually, suppose that there exist operators $d_1,\dots,d_k$ such
that
$$
 d_1\psi_1+\dots+d_k\psi_k=0.
$$
Let $d$ denote an matrix operator with operators $d_1,\dots,d_k$
in all rows. Then we have  $$
 (D(\lambda)+d)\psi=\lambda\psi,
$$
which contradicts the uniqueness of $D(\lambda)$. We show that if
$\psi_0\in M, \lambda\in A_Y,$ then $\lambda\psi_0\in M$. Let
$$
 \psi_0=d_1\psi_1+\dots+d_k\psi_k.
$$
By multiplying equality (2) from the left by a vector-row made up
by operators  $d_1,\dots,d_k$, we get  $$
 (d_1,\dots,d_k)D(\lambda)\psi=(d_1,\dots,d_k)\lambda\psi=\lambda(d_1,\dots,d_k)\psi
 =\lambda\psi_0,
$$
consequently, $\lambda\psi_0\in M$. Thus, apparently to every
commutative ring there connects a Baker-Akhiezer module which
satisfies the conditions 1,2 (see introduction).

In all examples the Baker-Akhiezer function has an essential
singularity on $Y$ and has the form $$
 \psi(x,P)=g(x,P)\exp(x_1F_1(P)+\dots+x_nF_n(P)),
$$
where $F_i(P), i=1,\dots,n$ are meromorphic  (in the general case
many-valued functions) on $X$ with pole on $Y$,
$$
 g(x,p)=(g_1(x,P),\dots,g_k(x,P))^{\top},
$$
 $g_i(x,P)$ are
meromorphic sections of the line bundle $L$ on $U\times X$, $U$ is
an open subset in ${\mathbb C}^n$ with pole on $\tilde{Y}=U\times
Y.$ The line bundle $L$ possesses connections
$$
 \nabla_i=\partial_{x_i}+F_i(P).
$$
It is clear that $\nabla_j g_i(x,P)$ is a section of bundle
$L\otimes [\tilde{Y}]^s, s>0$. In this way, we represent module
$M$ in the form $M=\cup_{m=1}^{\infty}M(m),$ where $M(m)$ consists
of functions
$$
 M(m)=\{\tilde{g}(x,P)\exp(x_1F_1(P)+\dots+x_nF_n(P)),\
 \tilde{g}(x,P)\in {\rm H^0}(L\otimes [\tilde{Y}]^m)\}.
$$
What is more, mappings
$$
 \partial_{x_j}:M(m)\rightarrow M(m+1)
$$
hold.

Further, for simplicity we assume that $n=2$.

First, we consider the case $k=1$. Let the ${\cal D}_2$-module $M$
be generated by the function $\psi$. We can assume that $\psi\in
M(1)$. Since the rank of the ${\cal O}$-module of differential
operators is of an order no higher than $m-1$ equal to
$\frac{m(m+1)}{2}$, by virtue of freeness of $M$ we have the
obvious equality
$$
 {\rm rank}_{\cal O}M(m)=\frac{m(m+1)}{2}.
$$
Since for any meromorphic function $\lambda$ with pole of order
$m$ on $Y$ there should be a unique differential operator
$D(\lambda)$ of order $m$ such that equality (2) is satisfied, so
in the general case equality
$$
 {\rm rank}_{\cal O}{\rm H}^0(U\times X, L\otimes
 [\tilde{Y}]^m)=\frac{m(m+1)}{2}
$$
should be fulfilled.  In this way, for the theory of commuting two
variable differential operators with scalar coefficients, the
following problem is important:

{\it Classification of algebraic manifolds $X$ of dimension 2,
such that
$$
 {\rm dim}{\rm H}^0(X, E\otimes
 [Y]^m)=\frac{m(m+1)}{2},\ m>0,
$$
where $E$ is a certain line bundle over $X$.}

Let $k=2$, $\psi_1,\psi_2$ be the basis in the ${\cal D}_2$-module
$M$. We can assume that $\psi_1\in M(1)$, $\psi_2\in M(m_0),
m_0\geq 1$. Then, by virtue of freeness of $M$ we have
$$
 {\rm rank}_{\cal
 O}M(m)=\frac{m(m+1)}{2}+\frac{(m-m_0+1)(m-m_0+2)}{2},\ m>m_0.
$$
Analogically, in order to find commuting two variable differential
operators with matrix coefficients of size $2\times 2$, the
following problem is important: $2\times 2$:

 {\it Classification of algebraic manifolds $X$ of dimension 2 such that
$$
 {\rm dim}{\rm H}^0(X, E\otimes
 [Y]^m)=\frac{m(m+1)}{2}+\frac{(m-m_0+1)(m-m_0+2)}{2},
 \eqno{(3)}
$$
$m>m_0$. }

Exactly in the same way we can examine the general case for
arbitrary $n$ and $k$.

\noindent {\bf Example } (A. Nakayashiki \cite{N1}). Let
$X^g={\mathbb C}^g/\{{\mathbb Z}^g+\Omega{\mathbb Z}^g\}$ be a
principle polarized Abelian  variety, with $\Omega$ being a
symmetric complex matrix with a positively defined imaginary part.
As subset $Y\subset X^g$ we consider a theta-divisor given by the
zeroes of a theta function that is defined by the absolutely
convergent series
$$
\theta(z)=\sum_{n\in{\mathbb Z}^g}\exp(\pi i\langle m,\Omega
m\rangle +2\pi i\langle m,z\rangle),\ z\in{\mathbb C}^g.
$$
Assume that the theta divisor is a non-singular subvariety. The
Baker-Akhie\-zer module has the form $M=\cup_{m=1}^{\infty}M(m),$
$$
M(m)= \left\{ g(z,x) \exp
\left(-x_1\partial_{z_1}\log\theta(z)\dots-x_g\partial_{z_g}\log\theta(z)\right)\!
 \right \},
$$
where $f(x,z)$ is a meromorphic function on ${\mathbb C}^g$ with a
pole of order not larger than $m$ on the theta-divisor, and also
periodic:
$$
 g(x,z+\Omega m+n)=\exp(-2 \pi i\langle m,c+x\rangle) g(x,z),\
 m,n\in{\mathbb Z}^g,
$$
$c\in{\mathbb C}^g\backslash \{0\}$ being a fixed point

\begin{itemize}
\item
$M$ is a free ${\cal D}_g$-module of rank $g!$.
\end{itemize}
Besides,
$$
 {\rm rank}_{\cal
 O}M(m)=m^g.
$$
In this way, for $g=2$ in formula (3) we have $m_0=2$.

\noindent {\bf Remark}. {\it Speaking more precisely, the
conditions formulated on increasing of ${\rm dim}{\rm H}^0(X,
 E\otimes [Y]^m)$
are necessary for the ${\rm gr}{\cal D}$-module ${\rm gr} M$ to be
free, where the graduation is respectively induced by the degree
of the operator and the order of pole on $Y$ (see details in
\cite{N1}). Maybe it is possible that free ${\cal D}$ modules $M$
exists, but ${\rm gr}{\cal D}$-modules ${\rm gr} M$ at the same
time are not free.}

\subsection{The Baker-Akhiezer module corresponding to the manifold
$\Gamma$} Now turn to our construction. The form $f(P)$ (see
introduction) is the section of the line bundle $\tilde{E}$ on
${\mathbb C}P^1\times {\mathbb C}P^1$, if identity (1) holds,  it
can be interpreted as a section of the line bundle on $\Gamma$
derived from $\tilde{E}$ by identification of the fibres over
double points. Let's introduce two forms
$$
 f_i(P)=\alpha_i z_1z_2+\beta_i z_1w_2+\gamma_i z_2w_1+\delta_i
 w_1w_2,\ \alpha_i,\beta_i,\gamma_i,\delta_i\in{\mathbb C},\ i=1,2,
$$
such that for $(t_1:t_2)\in{\mathbb C}P^1$  identity
$$
 \frac{f_i(a_1:b_1,t_1:t_2)}{f(a_1:b_1,t_1:t_2)}-
 \frac{f_i(t_1:t_2,a_2:b_2)}{f(t_1:t_2,a_2:b_2)}-c_i=0,\
 c_i\in {\mathbb C}. \eqno{(4)}
$$
holds. The dimension of spaces of such forms equals 3. By virtue
of (1) identity (4) is equivalent to
$$
 f_i(a_1:b_1,t_1:t_2)-A f_i(t_1:t_2,a_2:b_2)-A c_i
 f(t_1:t_2,a_2:b_2)=0.
$$

Chose $f_1$ and $f_2$ such $f_1,f_2$ and $f$ are linearly
independent (any other form that satisfies condition (4) with a
certain constant $c_i$ is a linear combination of $f_1$, $f_2$ and
$f$). Let
$$
 F_1(P)=\frac{f_1(P)}{f(P)},\ F_2(P)=\frac{f_2(P)}{f(P)}.
$$
Let $M(n)$ denote a set of functions of the form
$$
 M(n)=\left\{\psi=\frac{\tilde{f}(P,x,y)}{f^n(P)}
 \exp\left(x F_1(P)+y F_2(P) \right)\right\},
$$
for which the identity
$$
 \psi(a_1:b_1,t_1:t_2,x,y)-\Lambda\psi(t_1:t_2,a_2:b_2,x,y)=0,\
\Lambda\in{\mathbb C},\eqno{(5)}
$$
is fulfilled for all $(t_1:t_2)\in{\mathbb C}P^1$, where
$\tilde{f}$ is a form of the following kind:
$$
 \tilde{f}=\sum_{k,l=0}^nz_1^kw_1^{n-k}z_2^lw_2^{n-l}f_{kl}(x,y).\eqno{(6)}
$$
The set of functions from $M(n)$ forms a module above a ring of
analytic functions ${\cal O}$ on variables $x,y$ in the
nighbourhood of $0\in {\mathbb C}^2$. The rank of module $M(n)$
above ${\cal O}$ (dimension of the space of functions from $M(n)$
for fixed $x,y$) is equal to
$$
 {\rm rank}_{\cal O}M(n)=n(n+1).
$$

Actually, the dimension of the space of forms of kind (6) is equal
to $(n+1)^2$. Note that by virtue of (1) and (4) identity (5) is
equivalent to
$$
 \tilde{f}(a_1:b_1,t_1:t_2,x,y)-\tilde{f}(t_1:t_2,a_2:b_2,x,y)\Lambda A^n
 \exp(-xc_1-yc_2)=0.
$$
The last equality signifies that the coefficients of the
homogeneous polinomial on $t_1,t_2$ of $n$-th power standing in
the left-hand part equal zero. This imposes a  $(n+1)$ restriction
on the choice of the coefficients of form $\tilde{f}$. In this
way,  ${\rm rank}_{\cal O}M(n)=(n+1)^2-(n+1)=n(n+1)$ and,
consequently, in formula (3) for the manifold $\Gamma$ $m_0$=1.

Note that since $\Lambda$ does not depend on $x,y$ identity (5)
preserves its form at differentiation of $\psi$ by $x$ $y$.
Consequently, we have two mappings
$$
 \partial_x:M(n)\rightarrow M(n+1),\ \partial_y:M(n)\rightarrow
 M(n+1).
$$
In this way, the structure of the Baker-Akhiezer module above the
ring of differential operators
$$
 {\cal D}={\cal O}[\partial_x,\partial_y].
$$
is given on the set
$$
 M=\cup_{n=1}^{\infty}M(n)
$$
\section{Proof of theorem 1}
\subsection{Proof of freeness of the Baker-Akhiezer module}

Choose  two functions $\psi_1$ and $\psi_2$ in $M(1)$
$$
 \psi_i=\frac{h_i(P,x,y)}{f(P)}
 \exp\left(xF_1(P)+yF_2(P)\right),
$$
which are independent over ${\cal O}$ where
$$
 h_i(P,x,y)=a_i(x,y)z_1z_2+b_i(x,y)z_1w_2+c_i(x,y)w_1z_2+d_i(x,y)w_1w_2,\
 i=1,2.
$$
Functions $h_i$ satisfy the identity
$$
 h_i(a_1:b_1,t_1:t_2,x,y)-h_i(t_1:t_2,a_2:b_2,x,y)\Lambda A
 \exp(-xc_1-yc_2)=0.
$$

Let $P_1$ and $P_2$ denote the intersection points of the curves
given by equations $f_1(P)=0$ and $f(P)=0$, and by $Q_1$ and $Q_2$
the intersection points of the curves given by equations
$f_2(P)=0$ and $f(P)=0$. For their bulkiness, we do not cite the
explicit formulae for the point coordinates. $P_1,P_2$ and
$Q_1,Q_2$. By small movements of $c_1$ and $c_2$ we can achieve
these points to be mutually distinct.

The following theorem holds:

\begin{theorem}
For any function $\varphi\in M$ there exist two unique
differential operators $d_1,d_2\in {\cal D}$ such that
$$
 d_1\psi_1+d_2\psi_2=\varphi,
$$
i.e. $M$ is a free module over the ring of differential operators
${\cal D}$ of rank two, generated by functions $\psi_1,\psi_2$.
\end{theorem}

Theorem 1 follows directly from theorem 2.

Let $N$ denote the module over ${\cal D}$, generated by the
functions $\psi_1,\psi_2$
$$
 N=\{d_1\psi_1+d_2\psi_2,d_1,d_2\in {\cal D}\}.
$$
Theorem 2 follows from lemma 1 and lemma 2. In lemma 1 we show
that the ${\cal D}$-module $N$ is free of rank 2, and in lemma 2
we show that ${\cal D}$-moduli $M$ and $N$ coincide.

The following lemma holds:

\begin{lemma}
The module $N$ is free over the ring of differential operators
${\cal D},$ generated by functions $\psi_1,\psi_2$.
\end{lemma}

{\sl Proof of lemma 1.} Suppose that this module is not free, i.e.
there exist two differential operators $d_1,d_2\in D$ of order $n$
and $k$
$$
 d_1=\alpha_n(x,y)\partial_x^n+\alpha_{n-1}(x,y)\partial_x^{n-1}\partial_y+\dots
 +\alpha_0(x,y)\partial_y^n+\dots,
$$
$$
 d_2=\beta_k(x,y)\partial_x^k+\beta_{k-1}(x,y)\partial_x^{k-1}\partial_y+\dots+
 \beta_0(x,y)\partial_y^k+\dots,
$$
such that
$$
 d_1\psi_1+d_2\psi_2=0.\eqno{(7)}
$$
First, consider the case of $n\ne k$. For sake of definiteness,
let  $n>k$. Divide equality (7) by
$\exp\left(xF_1(P)+yF_2(P)\right)$, multiply by $f^{n+1}(P)$ and
after this contrain the resulting equality to a curve $f(P)=0$.
Get
$$
 h_1(P,x,y)(\alpha_n(x,y)f_1^n(P)+\alpha_{n-1}(x,y)f_1^{n-1}(P)f_2+\dots+
$$
$$
 \alpha_0(x,y)f_2^n(P))=0.\eqno{(8)}
$$
Substitute the resulting equality with point $P_1$
$$
 h_1(P_1,x,y)\alpha_0(x,y)f_2^n(P_1)=0.
$$
Direct verification shows that function $h_1(P_1,x,y)$ is not
identically equal to 0,  consequently, $\alpha_0(x,y)=0.$ Divide
(8) by $f_1$ and substitute again point $P_1$
$$
 h_1(P_1,x,y)\alpha_1(x,y)f_2^{n-1}(P_1)=0,
$$
consequently, $\alpha_1=0$. Analogically, we can show that $$
 \alpha_2=\dots=\alpha_n=0.
$$
Now consider the case of $k=n$. For $P\in\{f(P)=0\}$ in place of
(8) get
$$
 h_1(P,x,y)(\alpha_n(x,y)f_1^n(P)+\alpha_{n-1}(x,y)f_1^{n-1}(P)f_2+\dots
 +\alpha_0(x,y)f_2^n(P))+
$$
$$
 h_2(P,x,y)(\beta_n(x,y)f_1^n(P)+\beta_{n-1}(x,y)f_1^{n-1}(P)f_2+
 \dots+\beta_0(x,y)f_2^n(P)).\eqno{(9)}
$$
In (9) replace points $P=P_1$ and $P=P_2$. From this
$$
 h_1(P_1,x,y)\alpha_0(x,y)f_2^n(P_1)+h_2(P_1,x,y)\beta_0(x,y)f_2^n(P_1)=0,
$$
$$
 h_1(P_2,x,y)\alpha_0(x,y)F_2^n(P_2)+h_2(P_2,x,y)\beta_0(x,y)f_2^n(P_2)=0.
$$
Consequently,
$$
 \frac{h_1(P_1,x,y)}{h_2(P_1,x,y)}=
 \frac{h_1(P_2,x,y)}{h_2(P_2,x,y)}=\frac{\beta_0(x,y)}{\alpha_0(x,y)}.
$$
By direct verification find that
$$
 \frac{h_1(P_1,x,y)}{h_2(P_1,x,y)}\ne\frac{h_1(P_2,x,y)}{h_2(P_2,x,y)}\eqno(10)
$$
(here, for their bulkiness we do not cite explicit formulae for
$\frac{h_1(P_i,x,y)}{h_2(P_i,x,y)}$. For simplicity we do this for
a concrete example (see below)). Consequently,
$\alpha_0=\beta_0=0$. Dividing (9) by $f_1$ yields
$\alpha_1=\beta_1=0$. Analogically
$$
 \alpha_2=\dots=\alpha_n=\beta_2=\dots=\beta_n=0.
$$
Consequently, there exist no such operators $d_1$ and $d_2$. Lemma
1 is proven

The following holds:
\begin{lemma}
Modules $M$ and $N$ coincide.
\end{lemma}
{\sl Proof of Lemma 2 } Let $N(n)$ denote the following subset in
$N$
$$
 N(n)=\{d_1\psi_1+d_2\psi_2,\ d_1,d_2\in {\cal D},\  {\rm ord}\ d_1,{\rm ord}\ d_2\leq
 n-1\}.
$$
Since the ${\cal D}$-module $N$ is free,
$$
 {\rm rank}_{\cal O}N(n)=2\ {\rm rank}_{\cal O}\{d\psi_i,\ d\in {\cal D},
 \ {\rm ord}\ d\leq n-1\}=n(n+1).
$$
Consequently,
$$
 {\rm rank}_{\cal O}M(n)={\rm rank}_{\cal O}N(n).
$$
By virtue of the obvious inclusion $$
 N(n)\subset M(n),
$$
get
$$
 M=N.
$$
Thus, Lemma 2 is proven, jointly with theorem 2.

\subsection{Explicit formulas for commuting differential operators}

In this section, we give an example of commuting differential
operators and common vector eigen-functions which are parametrized
by the spectral manifold $\Gamma$.

Let points $p_1, p_2\in {\mathbb C}P^1$ have the following
coordinates
$$
 p_1=(1:0),\
 p_2=(0:1).
$$
We introduce three forms
$$
 f(P)=z_1z_2+z_1w_2+w_1w_2,
$$
$$
 f_1(P)=z_1z_2+2 w_1z_2-w_1w_2,
$$
$$
 f_2(P)=-z_1z_2+2w_1z_2+w_1w_2.
$$
By direct verification find that form $\theta(P)$ satisfies
identification (1) for $A=1$, and forms $f_1(P)$ and $f_2(P)$
satisfy identity (4), with $c_1=1$ and $c_2=-1$, respectively.

Curves $f(P)=0$ and $f_1(P)=0$ intersect in points $$
 P_1=(-2-\sqrt{2}:1,-\frac{1}{\sqrt{2}}:1),\
 P_2=(-2+\sqrt{2}:1,\frac{1}{\sqrt{2}}:1),
$$
and curves $f(P)=0$ and $f_2(P)=0$ --- in points
$$
 Q_1=(-\sqrt{2}:1,-1+\frac{1}{\sqrt{2}}:1),\
 Q_2=(\sqrt{2}:1,-1-\frac{1}{\sqrt{2}}:1),
$$

Take base $\psi_1,\psi_2$ in the ${\cal D}$-module $M$
$$
 \psi_1=\frac{w_1z_2}{f(P)}\exp\left(xF_1(P)+yF_2(P)\right),
$$
$$
 \psi_2=\frac{z_1z_2e^{y-x}+z_1w_2+w_1w_2e^{x-y}}{f(P)}\exp\left(xF_1(P)+yF_2(P)\right).
$$
Then,
$$
 \frac{h_1(P_1,x,y)}{h_2(P_1,x,y)}=-\frac{e^{x+y}}{\sqrt{2}(e^y-e^x)(-e^x+(1+\sqrt{2})e^y)},
$$
$$
 \frac{h_1(P_2,x,y)}{h_2(P_2,x,y)}=-\frac{e^{x+y}}{\sqrt{2}(e^y-e^x)(e^x+(-1+\sqrt{2})e^y)},
$$
thus, inequality (10) holds.

The four most simple meromorphic functions on $\Gamma$ with poles
on the curve $f(P)=0$ have the form
$$
  \lambda_1=\frac{w_1z_2}{f(P)},\
  \lambda_2=\frac{z_1w_1z_2^2}{f(P)^2},  \
  \lambda_3=\frac{z_1z_2w_1w_2}{f(P)^2},\
  \lambda_4=\frac{z_1w_1w_2^2+z_1^2z_2w_2}{f(P)^2}.
$$
The pairwise commutating operators corresponding to this function
have the form
$$
 D(\lambda_1)=
  \left(
  \begin{array}{cc}
    \frac{1}{4}(\partial_x+\partial_y) &  0\\
   0   &  \frac{1}{4}(\partial_x+\partial_y)\\
  \end{array}\right).
$$
$$
 [D(\lambda_2)]_{11}=\frac{e^x}{8(e^x-e^y)}(\partial_x^2-\partial_y^2)-
 \frac{e^{x+y}}{4(e^x-e^y)^2}(\partial_x+\partial_y),
$$
$$
 [D(\lambda_2)]_{12}=\frac{e^{x+y}}{16(e^x-e^y)^2}(\partial_x+\partial_y)^2,
$$
$$
 [D(\lambda_2)]_{21}=\frac{1}{8}(e^{y-x}-e^{x-y}-2)\partial_x^2+
 \frac{1}{8}(e^{x-y}-e^{y-x}-2)\partial_y^2+\frac{1}{2}\partial_x\partial_y+
$$
$$
 \frac{e^x+e^{2x-y}+5e^y-e^{2y-x}}{4(e^x-e^y)}\partial_x+
 \frac{3e^x-e^{2x-y}+3e^y+e^{2y-x}}{4(e^y-e^x)}\partial_y-
 \frac{e^y(2e^x+e^y}{(e^x-e^y)^2)},
$$
$$
 [D(\lambda_2)]_{22}=\frac{e^x}{8(e^y-e^x)}\partial_x^2-\frac{1}{4}\partial_x\partial_y+
 \frac{e^x-2e^y}{8(e^y-e^x)}\partial_y^2+
 \frac{e^y(2e^x+e^y)}{8(e^x-e^y)^2}(\partial_x+\partial_y).
$$
The operator corresponding to the function $\lambda_3$ has the
form
$$
 [D(\lambda_3)]_{11}=\frac{(e^x+e^y}{8(e^y-e^x)}(\partial_x^2-\partial_y^2)+
 \frac{(e^{2x}+e^{2y}}{4(e^y-e^x)^2}(\partial_x+\partial_y),
$$
$$
 [D(\lambda_3)]_{12}=\frac{e^{x+y}}{8(e^y-e^x)^2}(\partial_x-\partial_y)^2,
$$
$$
 [D(\lambda_3)]_{21}=\frac{1}{4}(2+e^{x-y}-e^{y-x})\partial_x^2-\partial_x\partial_y+\frac{1}{4}(2-e^{x-y}+e^{y-x})\partial_y^2+
$$
$$
 \frac{2e^x+e^{2x-y}+4e^y-e^{2y-x}}{2(e^y-e^x)}\partial_x+
 \frac{4e^x-e^{2x-y}+2e^y+e^{2y-x}}{2(e^x-e^y)}\partial_y+
 \frac{e^{2x}+e^{2y}+4e^{x+y}}{(e^x-e^y)^2},
$$
$$
 [D(\lambda_3)]_{22}=\frac{3e^x-e^y}{8(e^x-e^y)}\partial_x^2+\frac{1}{2}\partial_x\partial_y+\frac{e^x-3e^y}{8(e^x-e^y)}\partial_y^2-
 \frac{3e^{x+y}}{2(e^y-e^x)^2}(\partial_x+\partial_y).
$$
The operator corresponding to the function $\lambda_4$ has the
form
$$
 [D(\lambda_4)]_{11}=\frac{e^x+3e^y}{4(e^x-e^y)}\partial_x^2+\frac{1}{2}\partial_x\partial_y-\frac{3e^x+e^y}{4(e^x-e^y)}\partial_y^2-
$$
$$
 \frac{e^{2x}+3e^{2y}}{2(e^x-e^y)^2}\partial_x-\frac{3e^{2x}+e^{2y}}{2(e^x-e^y)^2}\partial_y-
 \frac{2e^{x+y}}{(e^y-e^x)^2},
$$
$$
 [D(\lambda_4)]_{12}=\frac{e^{x+y}}{2(e^y-e^x)^2}((\partial_x+\partial_y)^2+\partial_x+\partial_y),
$$
$$
 [D(\lambda_4)]_{21}=(e^{y-x}-e^{x-y}-2)\partial_x^2+4\partial_x\partial_y+(e^{x-y}-e^{y-x}-2)\partial_y^2+
$$
$$
 \frac{5e^x+e^{2x-y}+9e^y-3e^{2y-x}}{e^x-e^y}\partial_x+
 \frac{9e^x-3e^{2x-y}+5e^y+e^{2y-x}}{e^x-e^y}\partial_y+
$$
$$
 \frac{2e^{-x-y}(e^{4x}+e^{4y}-6e^{2(x+y)}-4e^{3x+y}-4e^{x+3y})}{(e^x-e^y)^2},
$$
$$
 [D(\lambda_4)]_{22}=\frac{7e^x-3e^y}{4(e^y-e^x)}\partial_x^2-\frac{3}{2}\partial_x\partial_y+\frac{3e^x-7e^y}{4(e^y-e^x)}\partial_y^2-
$$
$$
\frac{e^{2x}+3e^{2y}-16e^{x+y}}{2(e^x-e^y)^2}\partial_x-
 \frac{3e^{2x}+e^{2y}-16e^{x+y}}{2(e^x-e^y)^2}\partial_y.
$$

$$
$$

\noindent Sobolev Institute of Mathematics,

\noindent Novosibirsk State University,

\noindent {\it E-mail address:} mironov@math.nsc.ru

\end{document}